\documentclass[runningheads]{llncs}
\usepackage{listings}
\usepackage{xcolor}
\usepackage{xspace}
\usepackage{trfrac}
\usepackage{mathtools}
\usepackage{marvosym}
\usepackage{amssymb}
\usepackage{amsmath}
\usepackage{cite}
\usepackage{relsize}

\usepackage[hidelinks]{hyperref}

\definecolor{bluekeywords}{rgb}{0.13,0.13,1}
\definecolor{greencomments}{rgb}{0,0.5,0}
\definecolor{redstrings}{rgb}{0.9,0,0}

\lstdefinestyle{codeStyle}{
    breakatwhitespace=false,
    breaklines=true,
    captionpos=b,
    keepspaces=true,
    numbers=left,
    numbersep=5pt,
    showspaces=false,
    showstringspaces=false,
    showtabs=false,
    commentstyle=\color{greencomments},
    keywordstyle=\color{bluekeywords},
    stringstyle=\color{redstrings},
    tabsize=2,
    basicstyle=\ttfamily,
    aboveskip=1em
}

\newenvironment{Grammar}{$\begin{array}[t]{lcll}}{\end{array}$}
\newcommand\ProductionDef{%
  \Coloneqq\ 
  }
\newcommand{\Production}[3]{%
\!\!\!#1\ %
&%
\!\!\!\ProductionDef\!\!%
&%
#2%
&%
\!\!\mbox{{\small{#3}}}%
}
\newcommand{\MoreProduction}[2]{%
&&%
#1 
&%
\!\!\!\mbox{{\small{#2}}}%
}

\newcommand{\NonTerminal}[1]{\ensuremath{\mathit{#1}}\xspace}
\newcommand\Q\lstinline

\usepackage{letltxmacro}
\newcommand*{\SavedLstInline}{}
\LetLtxMacro\SavedLstInline\lstinline
\DeclareRobustCommand*{\lstinline}{%
  \ifmmode
    \let\SavedBGroup\bgroup
    \def\bgroup{%
      \let\bgroup\SavedBGroup
      \hbox\bgroup
    }%
  \fi
  \SavedLstInline
}

\begin{document}

\title{Smoothly Navigating between Functional Reactive Programming and Actors}
\titlerunning{Smoothly Navigating between FRP and Actors}

\author{Nick Webster\inst{1} \and Marco Servetto\inst{1}}

\institute{
    Victoria University of Wellington, Kelburn, Wellington, 6012, New Zealand
    \email{\{nick@nick.geek.nz, marco.servetto@ecs.vuw.ac.nz\}}
    \{\href{https://nick.geek.nz}{https://nick.geek.nz}, \href{http://ecs.victoria.ac.nz/Main/MarcoServetto}{http://ecs.victoria.ac.nz/Main/MarcoServetto}\}
}

\maketitle

\begin{abstract}
    We formally define an elegant multi-paradigm unification of Functional Reactive Programming,
    Actor Systems, and Object-Oriented Programming. This enables an intuitive form of declarative
    programming, harvesting the power of concurrency while maintaining safety.

    We use object and reference capabilities to highlight and tame imperative features:
    reference capabilities track aliasing and mutability, and object capabilities track I/O.
    Formally, our type system limits the scope, impact and interactions of impure code.
    \begin{itemize}
        \item Scope: Expressions whose input is pure will behave deterministically.
        \item Impact: Data-races and synchronisation issues are avoided. The only
        way for an actor to behave nondeterministically, is by mutating its state based on message delivery order.
        \item Interactions: Signals provide a functional boundary between
        imperative and functional code, preventing impure code from invalidating functional assumptions.
    \end{itemize}
    
\end{abstract}

\section{Introduction}
Parallel programming promises great performance improvements, but it is also a source of undesired nondeterministic behaviour.
Actor systems and FRP (Functional Reactive Programming) tame nondeterminism in different ways:
each actor sees the world sequentially, and processes a single message at a time. However, messages can be delivered in an unpredictable order.
Instead, pure FRP guarantees complete determinism; signals may be processed in various orders and in parallel, but immutability shields us from observing any parallelism. 

In 2019, Lohstroh et al. \cite{determActors} proposed a new actor system that uses \textit{reactors}. Reactors declare their inputs and outputs,
react to messages, and are connected with a \textit{composite} main function that builds the graph. Effectively, the system uses reactive programming
techniques to build actor systems. The `reactor network' can offer stronger guarantees for message delivery/processing order than traditional actor models. However the system does not enforce any properties on behaviour, like determinism
or the absence of data races.

In this paper we propose \textit{Featherweight Reactive Java} (FRJ), a way to blend FRP with actor systems in a minimal subset of Java, inspired by 
Featherweight Java (FJ) \cite{igarashi2001featherweight}. FRJ achieves this by using functional
reactive programming techniques, which offers all the benefits of the reactor model
and control over I/O and state mutation. FRJ fulfils the promise of Lohstroh et al.'s work by
allowing the declarative creation of truly deterministic actor systems.

In actor systems, actors have a list of messages that they process one at a time.
For simplicity, in our language every object can function as an actor, and thus in memory there is a list of messages near every object record.
FRJ's FRP is inspired by E-FRP's discrete signals, which are signals that update upon events occurring \cite{wan2002event,czaplicki2012elm}.
FRJ's signals are a possibly infinite sequence of messages. The messages contain values that have either been computed
or are being computed. The head is the most recent message; the tail is an expression that returns a new signal for the next message.
Any usages of a signal with a message that has yet to be computed, will wait for the computation to finish.
We implement signals via lists of expressions. The computation of each expression is deferred and the result
of the computation is a `message'.
The head and tail of the signal can be accessed using the conventional \Q{head(_)} and \Q{tail(_)} syntax.
Finally, we have a special syntax for lifted method calls:
\Q|a.@m(b,c)|, where \Q@b@ and \Q@c@ are signals. This syntax sends the actor
\Q@a@ a message causing the (asynchronous) computation of \Q@a.m(head(b),head(c))@
and then triggers \Q|a.@m(tail(b),tail(c))|; until either \Q@b@ or \Q@c@ terminates.

We can connect real world input output with our signals by using object capabilities.
We have an expressive type system based on reference capabilities, supporting two fundamental properties:
expressions that only use immutable references are deterministic, and parallelism can only induce nondeterminism if a mutable actor relies on the delivery order of messages.

\section{FRJ}
Consider the following class:
\begin{lstlisting}
class Person {
  method Int age() {return 24;} method Str name() {return "Bob";}
  method Str format(Str name,Int age) { return name+":"+age; }}
\end{lstlisting}

\noindent Using it, we could write the conventional method call \\*\Q@p.format(p.name(),p.age())@,\\* to compute a string once.
Using FRJ's lifted method calls, we can write the following in an FRP style:
\begin{lstlisting}
@Int ages = p.@age();
p.@format(p.@name(),ages);
\end{lstlisting}

This creates a signal with messages containing formatted names and ages. If we connected the behaviour of \Q@age()@ to the real world, we would see the formatted names change as Bob grows older.
We can also write code in the actor style, by sending individual events to \Q@p@; 
\Q|@[]| is the empty signal and 
the syntax \Q|@[  ;  ]| builds a signal manually.
\begin{lstlisting}
@Int ages=@[p.age();@[]]
a.@format(@[p.name();@[]],ages)
\end{lstlisting}

This code sends the messages, \Q@age@ and \Q@name@,
to the actor \Q@p@ one time. \Q@p@ replies with an \Q@Int@ and a \Q@Str@ message.
When both messages are handled, \Q@p@ receives a single message asking to produce the formatted string, parameterised over the name and the age.
The creation of messages inside signals is computed in parallel and execution is deferred. Thus, implementing a fork--join is trivial in FRJ:
\begin{lstlisting}
@Int part1=@[x.computePart1();@[]]
Int part2=x.computePart2();
return head(part1)+part2;
\end{lstlisting}

The fork-join works because the creation of \Q@part1@ does not block because its head is being
evaluated in parallel. The \Q@head(part1)@ call would block until the expression had been computed and the message was ready.

\subsection{Grammar}
\renewcommand\Cap{\NonTerminal{cap}} 
\newcommand\capability{\Q@capability@}
\newcommand\CD{\NonTerminal{CD}}
\newcommand\C{\NonTerminal{C}}
\newcommand\Cs{\NonTerminal{\overline{\C}}}
\newcommand\T{\NonTerminal{T}}
\newcommand\Ts{\NonTerminal{\overline{\T}}}
\newcommand\x{\NonTerminal{x}}
\newcommand\xs{\NonTerminal{\overline{x}}}
\renewcommand\v{\NonTerminal{v}}
\newcommand\vs{\NonTerminal{\overline{\v}}}
\newcommand\m{\NonTerminal{m}}
\newcommand\ms{\NonTerminal{\overline{\m}}}
\newcommand\f{\NonTerminal{f}}
\newcommand\fs{\NonTerminal{\overline{\f}}}
\newcommand\F{\NonTerminal{F}}
\newcommand\Fs{\NonTerminal{\overline{\F}}}
\newcommand\K{\NonTerminal{K}}
\newcommand\Ks{\NonTerminal{\overline{\K}}}
\newcommand\M{\Sigma}
\newcommand\Ms{\NonTerminal{\overline{\M}}}
\newcommand\MH{\NonTerminal{MH}}
\newcommand\MHs{\NonTerminal{\overline{\MH}}}
\newcommand\mdf{\NonTerminal{mdf}}
\newcommand\ctx{%
  \ensuremath{\mathcal{E}}%
}
\renewcommand\S{\NonTerminal{S}}
\newcommand\Ss{\NonTerminal{\overline{\S}}}
\renewcommand\L{\NonTerminal{L}}
\newcommand\e{\NonTerminal{e}}
\newcommand\es{\NonTerminal{\overline{\e}}}
\newcommand\hole{\NonTerminal{\square}}
\newcommand\Meth{\NonTerminal{M}}
\newcommand\Meths{\NonTerminal{\overline{\Meth}}}
\newcommand\Msg{\NonTerminal{\text{\Letter}}}
\newcommand\Msgs{\NonTerminal{\overline{\Msg}}}
\newcommand\Postal{\NonTerminal{\rho}}
\newcommand\Mem{\NonTerminal{\mu}}
\newcommand\class{\Q@class@}
\newcommand\method{\Q@method@}
\newcommand\mdfImm{\Q@imm@}
\newcommand\mdfMut{\Q@mut@}
\newcommand\mdfRead{\Q@read@}
\newcommand\mdfCapsule{\Q@capsule@}
\newcommand\oC{\ \Q@\{@}
\newcommand\cC{\ \Q@\}@}
\newcommand\oR{\ \Q@(@}
\newcommand\cR{\Q@)@}
\newcommand\oS{\Q@[@}
\newcommand\cS{\Q@]@}
\newcommand\semi{\!\Q@;@\!}
\newcommand\comma{\Q@,@}
\newcommand\sdot{\Q@.@}
\newcommand\colm{:}
\newcommand\at{\Q@\@@}
\newcommand\this{\Q@this@}
\newcommand\thisd{\Q@this.@}
\newcommand\return{\Q@return@}
\newcommand\mcall[1]{\Q@.@m\Q@(@#1\cR}
\newcommand\lmcall[1]{\Q@.@\at \m\Q@(@#1\cR}
\newcommand\classI[1]{C\Q@(@#1\Q@)@}
\newcommand\new[1]{\Q@new\ @\classI{#1}}
\newcommand\scon[3]{#1\oS#2\semi#3\cS}
\newcommand\sconE{\at\oS\, \cS}
\newcommand\head[1]{\Q@head(@#1\cR}
\newcommand\tail[1]{\Q@tail(@#1\cR}
\newcommand\memLV{\L}
\newcommand\Object{\Q@Object()@}
\newcommand\methTypes{\mathit{methTypes}}
\newcommand\validActor{\mathit{validActor}}
\newcommand\overrideOk{\mathit{overrideOk}}
\newcommand\fields{fields}
\newcommand\capOf{\mathit{capOf}}
\newcommand\envs{\Cap;\M;\Gamma \vdash}
\newcommand\Define{\Q@\#define@}
\newcommand\OK{\mathrm{OK}}
\newcommand\OKINC{\text{OK IN }\C}

\begin{Grammar}
\Production{\Cap}{\capability \mid \emptyset}{}\\
\Production{\CD}{
    \Cap\ \ \class\ \C\ \Q@implements@\ \Cs
    \oC\Fs\ \K\ \Meths \cC
    \mid \Q@interface@\ \C\ \Q@extends@\ \Cs  
    \oC \MH_1\semi\ldots\MH_n\semi\cC
}{}\\
\Production{\K}{
    \C\oR \T_1\,\x_1\ldots\T_n\,\x_n\cR
    \oC\thisd\x_1\Q@=@\x_1\semi\ldots
    \thisd\x_n\Q@=@\x_n\semi\cC
}{}\\
\Production{\F}{\T\,\f\!\semi}{}\\
\Production{\T}{\mdf\,\C \mid \Q|@|\T}{}\\
\Production{\MH}{\mdf\ \method\ \T\ \m\oR\T_1\ \x_1 \ldots \T_n\ \x_n \cR\,\semi}{}\\
\Production{\Meth}{\MH\ \oC\return\ \e\,\semi\cC}{}\\
\Production{\e}{
    \x
    \mid \e\mcall{\es}
    \mid \e\sdot\f
    \mid \e_1\sdot\f \Q@=@ \e_2
    \mid \new{\es}
    \mid \e\lmcall{\es}
    \mid \scon{\at}{\e\,}{\e'}
    \mid \sconE
    \mid \head{\e}
    \mid \tail{\e}
}{}\\
\Production{\v}{\memLV \mid \S \mid \oS\v\semi\S\cS}{}\\
\Production{\ctx}{
    \hole
    \mid \ctx\mcall{\es}
    \mid \v\mcall{\vs\, \ctx\, \es}
    \mid \ctx\lmcall{\es}
    \mid \v\lmcall{\vs\, \ctx\, \es}
    \mid \ctx\,\sdot\f
    \mid \ctx\,\sdot\f \Q@=@ \e
    \mid \v\sdot\f \Q@=@\, \ctx
}{}\\
\MoreProduction{
    \mid \new{\vs\, \ctx\, \es}
    \mid \head{\ctx}
    \mid \tail{\ctx}
}{}\\
\Production{\mdf}{\mdfImm \mid \mdfMut \mid \mdfCapsule \mid \mdfRead}{}\\
\Production{\Msg}{\scon{\S}{\e_1}{\e_2}}{}\\
\Production{\Mem}{\Postal_1 \ldots \Postal_n}{}\\
\Production{\Postal}{\L \mapsto \classI{\vs}\ \Msgs}{}
\end{Grammar}

\vspace{1em}

FRJ is a minimal OO language, where the class table contains classes or interfaces. Classes have methods, $\Meth$, fields, $\F$, and a conventional constructor, $\K$ initialising all of the fields. Interfaces provide conventional nominal subtyping, and for simplicity we do not offer any kind of subclassing. The language makes use of modifiers
to implement reference capabilities. The modifiers will be discussed alongside the typing rules because they are transparent to the reduction.
FRJ builds over the conventional small step reduction model where a pair $\Mem |\e $ is reduced into a new memory and a new expression.
The $\ctx$ nonterminal is the evaluation context for the reduction.
The memory is a map from object locations $\L$ to 
conventional object records. Additionally, every record also maintains a list of pending messages ($\Msgs$).
Values are conventional object locations $\L$, future signal values $\S$, and completed signal values $\oS\v\mid\S\cS\,$.
Types are class or interface names annotated with a capability modifier \mdf. The default modifier \Q@imm@ can be omitted for convenience.
Types for signals are annotated with \Q|@|. We also support higher-order signals, as \Q|@@T|.

In addition to conventional method calls, field accesses, field updates, and constructor calls, FRJ offers 
lifted method calls
$\e\,\lmcall{\es}$\,,
explicit signal construction
$\scon{\at}{\e\,}{\e'}$,
and the conventional  $\head{\e}$ and $\tail{\e}$  notation.
FRJ also offers the empty signal $\sconE$, which is a special signal that will not have any more messages in it.
The special variable $\this$ is implicitly provided as an argument to methods.

We support some imperative features like field updates to prove that we can tame imperative behaviour whilst retaining our guarantees.
If you want to tame a lion, you need to go into the lion's cage first. The more traditional functional approach of using recursion to keep state
can also be used, including within signals.
That is, programmers using FRJ can navigate between the functional and the imperative programming styles without losing the functional guarantees in the functional parts.
This means that a programmer wanting to use some in-place mutations in their code can do so privately, and a programmer using such code from a functional environment will not be able to observe the presence (or absence) of imperative features.
Moreover, a programmer starting from an imperative setting (even with direct connection to IO) can call functional code to enjoy functional properties on such calls.

\subsection{Well-Formedness}
Using the auxiliary notation, our well-formedness rules are as follows:
\begin{itemize}
\item $\sconE\ $ is not in $domS(\Mem)$ (defined below).
    \item All classes and interfaces are uniquely named.
    \item All methods in a given class are uniquely named.
    \item All fields in a given class are uniquely named.
    \item All parameters in a given method are uniquely named and are not called \this\,.
    \item A \Q@capsule@ method parameter can be used zero or one times in the method body
    \item All $\S$ labelling a $\Msg$ inside the memory are unique.
    \item Fields can only have the type modifiers: \,\mdfImm\,\ or\ \,\mdfMut\,.
    \item Types containing \Q|@| must have the \Q@imm@ modifier.
    \item Classes can only implement interfaces.
    \item Interfaces can only extend other interfaces.
    \item $\Mem \mid e$ is well formed if all $L$ in $e$ are in $dom(\Mem)$ (defined below)
    and all $usedS(e)\, \cup\, usedS(\Mem)$ are in $domS(\Mem)$.
\end{itemize}

$dom(\Mem)$ is the conventionally defined set of all keys ($L$) in the map ($\Mem$).
$domS(\Mem)$ is the set of all $\S$ labelling a $\Msg$ inside the memory, and 
$usedS(\Mem)$ is defined as follows:
\begin{itemize}
    \item $usedS(\Mem) = usedS(\Mem,\Postal_1) \cup\ldots\cup usedS(\Mem,\Postal_n)$,\\* with $\Mem=\Postal_1\ldots\Postal_n$
    \item $usedS(\L \mapsto \classI{\v_1\ldots\v_k}\,\ \Msg_1 \ldots\Msg_n) =$\\* $usedS(\v_1) \cup\ldots\cup usedS(\v_k) \cup usedS(\Msg_1) \cup\ldots\cup usedS(\Msg_n)$
    \item $\S \in usedS(\scon{\S'}{\e_1}{\e_2\!})\ = usedS(\e_1) \cup usedS(\e_2)$
    \item $\S \in usedS(\e) $ if \S is a sub-expression of $\e$.
\end{itemize}

\subsection{Reduction Rules}
The shape of the reduction is: $\Mem \mid \e \to \Mem' \mid e'$.
We use $class(\C)$ to denote the class declaration ($\CD$) for the class $\C$
and $\fields(\C)$ to denote the list of the fields for the class $\C$.
Additionally, `$\_$' is used as a placeholder in the rules and can match any syntactic term.

\[
    \trfrac[(fAccess)]{
        \begin{trgather}
            \L \mapsto \classI{\v_1\ldots\v_n}\ \_\ \text{in}\ \mu\qquad
            \T_1\, \f_1\ldots\T_n\, \f_n = \fields(\C)
        \end{trgather}
    }{
        \Mem \mid L.f_i \to \Mem \mid v_i
    }
\]

\[
    \trfrac[(fUpdate)]{
        \begin{trgather}
            \_\ \T_0\, \f_0\ldots\T_n\, \f_n = \fields(\C)\quad
            \Postal_0 = \L \mapsto \C\oR\vs\, \v_0\ldots\v_n)\, \Msgs\quad
            \Postal_1 = \L \mapsto \C\oR\vs\, \v\, \v_1\ldots\v_n)\, \Msgs 
        \end{trgather}
    }{
        \begin{trgather}
            \Mem\,,\Postal_0 \mid \L.\f_0 = \v
                \to
            \Mem\,,\Postal_1 \mid \v
        \end{trgather}
    }
\]

\[
    \trfrac[(new)]{
    }{
        \begin{trgather}
            \Mem \mid \new{\v_1\ldots\v_n}\,
                \to
            \Mem,\, \L \mapsto \classI{\v_1\ldots\v_n}\, \Msgs \mid \L
        \end{trgather}
    }
\]

\[
    \trfrac[(mCall)]{
        \begin{trgather}
            \L \mapsto \classI{\vs}\, \Msgs\, \text{in}\ \Mem\quad
            \_\ \method\ \_\ \m\oR\_\ \x_1\ldots\_\ \x_n\cR\oC\return\ \e\,\semi\cC\ \text{in}\ class(\C)
        \end{trgather}
    }{
        \begin{trgather}
            \Mem \mid \L \mcall{\v_1\ldots\v_n}\ \to \Mem \mid \e[\this=\L
            \comma\, \x_1=\v_1
            \ldots\x_n=\v_n]
        \end{trgather}
    }
\]

\[
    \trfrac[(\ctx)]
    {\Mem \mid e \to \Mem \mid e'}
    {\Mem \mid \ctx[e] \to \Mem \mid \ctx[e']}
\]

\[
    \trfrac[(\ctx{}Head)]{
        \Mem \mid \e \to \Mem \mid \e'
    }{
        \begin{trgather}
            \Mem,\, \Postal\, \scon{\S}{\ctx[\e]}{\e_0}\ \mid \e_1
                \to
            \Mem,\, \Postal\, \scon{\S}{\ctx[\e']}{\e_0}\ \mid \e_1
        \end{trgather}
    }
\]

\[
    \trfrac[(\ctx{}Tail)]{
        \Mem \mid \e \to \Mem \mid \e'
    }{
        \begin{trgather}
            \Mem,\, \Postal\, \scon{\S}{\v\,}{\ctx[\e]}\ \mid \e_1
                \to
            \Mem,\, \Postal\, \scon{\S}{\v\,}{\ctx[\e']}\ \mid \e_1
        \end{trgather}
    }
\]

Field updates, field access, object construction and method call are standard.
Contextual rules (\ctx, \ctx{}Head, and \ctx{}Tail) guide the parallel reduction: (\ctx) allows us to reduce the main expression,
while (\ctx{}Head) reduces the last message of an object.
When the value is produced, rule (\ctx{}Tail) executes the expression creating the next stream node.
\textbf{}Note that since the memory $\Mem$ is a set, the rules can work on any $\Postal$ in $\Mem$.
The $\Postal$ non-terminal has a list of $\Msg$ at its end; thus by writing 
$\Mem,\, \Postal\, \scon{\S}{\e\,}{\e'}$ we are selecting the last message of an arbitrary object in memory.

\[
    \trfrac[(head)]{
    }{
        \Mem \mid \head{\oS\v\,\semi\S\cS} \to \Mem \mid \v
    }
\]

\[
    \trfrac[(tail)]{
    }{
        \Mem \mid \tail{\oS\v\,\semi\S\cS} \to \Mem \mid \S
    }
\]

\[
    \trfrac[(tailEmpty)]{
    }{
        \Mem \mid \tail{\sconE} \to \Mem \mid \sconE
    }
\]

\[
    \trfrac[(msgComplete)]{
    }{
        \begin{trgather}
            \Mem,\, \Postal\, \scon{\S}{\v\,}{\S'}\ \mid \e_1
                \to
            (\Mem,\, \Postal\, \mid \e_1)[\S=\oS\v\,\semi \S'\cS]
        \end{trgather}
    }
\]

\[
    \trfrac[(Empty)]{
    \begin{trgather}
    \text{either }\e=\ctx[\head{\sconE}]\\
    \text{or }\e=\v \text{ and } \e_0=\ctx[\head{\sconE}]
    \end{trgather}
    }{
        \begin{trgather}
            \Mem,\, \Postal\, \scon{\S}{\e\,}{\e_0}\ \mid \e_1
                \to
            (\Mem,\, \Postal\, \mid \e_1)[\S=\sconE]
        \end{trgather}
    }
\]

The rule (head) is conventional and simply reduces
to the current value of a completed signal. When the expression is well typed,
the tail of a message is expected to be a signal that will eventually contain
the next value, rule (tail) can be used to get that continuation signal.

When a message has been completely computed, rule (msgComplete) removes the message from the memory, and replaces all of the references to $\S$ with $\oS\v\semi\S\cS$. So, while $\head{\S}$ will cause the reduction to get stuck,
$\head{\oS\v\semi\S\cS}$ can reduce. Therefore, the rule (msgComplete) enables a form of synchronisation
between the messages and their consumers.

When the message execution tries to access the head of the empty signal,
rule (Empty) terminates the signal, removing the message $\S$ and by replacing all occurrences of $\S$ with $\sconE$.

\[
    \trfrac[(explicitS)]{
    }{
        \begin{trgather}
            \Mem \mid \scon{\at}{\e_1}{\e_2}
                \to
            \Mem\comma\, \L \mapsto \Object\ \scon{\S}{\e_1}{\e_2} \mid \S
        \end{trgather}
    }
\]

\[
    \trfrac[(liftS)]{
        \begin{trgather}
            \e_0 = \L\lmcall{\v_1\ldots\v_n}\\
            \e_1 = \L\mcall{\head{\v_1}\ldots\head{\v_n}}\qquad
            \e_2 = \L\lmcall{\tail{\v_1}\ldots\tail{\v_n}}\qquad
        \end{trgather}
    }{
        \mu, \L \mapsto \classI{\vs}\ \Msgs \mid e_0
            \to
        \mu, \L \mapsto \classI{\vs}\ \scon{\S}{\e_1}{\e_2\!}\,\Msgs\mid \S
    }
\]

This group of two rules (explicitS, and liftS) deals with
the creation of signals.

For the creation of signals, the rule (explicitS) reduces signal constructors into a message ($\Msg$) and
places it on a new empty actor. The signal constructor expression is then replaced with
the fresh signal ($\S$) that was just associated with the message.

The alternative way to create signals in FRJ is through lifting methods.
liftS reduces lifted method calls by creating a $\Msg$ that gets placed onto the receiver
containing a head of the traditional method call with arguments of the head of all of its inputs.
The tail of this new $\Msg$ will be the same lifted method call, but with the tail of all of the inputs
as the inputs for the new lifted call. Effectively, the method now \textit{reacts} to its inputs.

\[
    \trfrac[(garbage)]{
    }{
        \Mem,\Mem' \mid \e \to \Mem \mid \e
    }
\]
Finally, the rule (garbage) gets rid of the part of memory that is unreachable starting from the main expression.
Note that we cannot arbitrarily split the memory. We can only split it in such a way that the resulting $\Mem \mid \e$ is well
formed.
An important consequence of our garbage collection rule is that messages can be collected too, even during their computation.
However, due to our well-formedness rules, messages can only be collected if the receiver actor object is collected, and an object can only be collected if there are no other references to its address and to any of the $\S$ in its mailbox.

\subsection{Reference capabilities}
Parallel computation is inherently part of FRP and actor systems. FRJ uses reference capabilities to tame the nondeterminism that would otherwise arise from aliasing and mutability.
FRJ supports the three traditional reference capabilities: \Q@imm@, deeply immutable (the default);
\Q@mut@ mutable and \Q@read@, the common supertype of both \Q@imm@ and \Q@mut@.
In addition, FRJ supports \Q@capsule@; a reference that dominates its $\mathrm{ROG}_{\mdfMut}$ (reachable object graph) \cite{giannini2019flexible}.
In OO languages, $\mathrm{ROG}(\L) = \overline{\L}$ is all the locations transitively reachable from the fields of $\L$.
With reference capabilities, mutable $\mathrm{ROG}_{\mdfMut}(\L) = \overline{\L}$ is all the locations transitively reachable from $\L$ only following mutable fields.

Assuming a traditional \Q@Person@ class, the following is an example of reference capabilities:
\begin{lstlisting}
mut Person mP = new Person("Bob", 24);
imm Person iP = new Person("Bob", 24);
read Person rP = mP;
mP.setAge(25); //ok, now rP.getAge() == 25
iP.setAge(25); //type error
rP.setAge(25); //type error
rP = iP; //ok, read is supertype of imm/mut
\end{lstlisting}

Note how the same object may be pointed at the same time by multiple references with different modifiers.
Capsule references can be obtained when the aliasing is under control, and can be used 
to create immutable references.
Capsule references can be used to create immutable references from non-immutable objects. Capsule references
can only exist in expressions, and the whole mutable object graph reachable from a capsule reference can only
be reached from that specific capsule reference. In this way, the capsule reference is the sole access point
to a group of mutable objects.
Reference capabilities have the
following subtype relation:
\begin{itemize}
    \item $\mdfCapsule\,\ \leq \mdf$
    \item $\mdf \leq \,\mdfRead$
\end{itemize}

Thus, all of the reference capabilities are subtypes of \Q@capsule@ and supertypes of \Q@read@.
\Q@mut@ and \Q@imm@ are not comparable to each other.

The main advantage of reference capability over older forms of aliasing control \cite{boyland2003checking, hogg1991islands}, is that references can be promoted/recovered to a subtype when the right conditions arise.
In this work we rely only on \textit{multiple method types}:
\[
\trfrac[]{
\begin{trgather}
\mdf\ \method\ \T\ \m\oR\T_1\ \x_1\ldots\T_n\ \x_n\cR\ \_\ \in\,class(\C)\\
\T_0=\mdf'\ \C\qquad
\mdf' \leq \mdf
\end{trgather}
}{
\begin{trgather}
\methTypes(\T_0, \m)=\{\T_0\ldots\T_n \mapsto \T,\\
(\T_0\ldots\T_n \mapsto \T)[\,\mdfMut\,=\,\mdfCapsule \,],\\
(\T_0\ldots\T_n \mapsto \T)[\,\mdfMut\,=\,\mdfCapsule\,, \,\mdfRead\,=\,\mdfImm\,]\}
\end{trgather}
}
\]

Where the notation $[\,\mdfMut\,=\,\mdfCapsule \,],$
replaces all of the $\mdfMut\,$ modifiers with $\mdfCapsule\,$.

For example,
the following code is correct:
\begin{lstlisting}
class Box {
  mut F f;
  Box(mut F f){ this.f = f; }
  read method read F f(){ return this.f; }
  }
class MakeBox{
  method mut Box of(mut F f){ return new Box(f); }
  }
...  
capsule F f = .. //we have a capsule f
capsule Box b = new MakeBox().of(f);
imm Box immB = b;
imm F immF = immB.f();
\end{lstlisting}
On line 7, method \Q@of(f)@ was declared taking an \Q@imm@ receiver and a \Q@mut@ parameter, and returning a \Q@mut@, but when called with a \Q@capsule@ parameter (line 11), we can promote the result to \Q@capsule@.
On line 4, method \Q@f()@ was declared taking a \Q@read@ receiver and returning a \Q@read@,
but when called with an \Q@imm@ receiver (line 13) we can promote the result to \Q@imm@.
\subsection{Object capabilities}
An object capability is an object whose methods can do privileged operations.
While reference capabilities keep mutability and aliasing under control, we rely on object capabilities \cite{melicher_et_al:LIPIcs:2017:7270} to tame I/O.
Our reduction rules do not model I/O directly, but we assume predefined \Q@capability@ classes containing \Q@mut@ methods doing all of the desired I/O interactions.
Since only \Q@mut@ methods of \Q@capability@ classes can do nondeterministic I/O, we keep I/O under control by allowing only
the \Q@main@ and \Q@mut@ methods of capability classes to create instances of capability classes \cite{finifter2008joee}.
In this way, any method that only takes immutable objects as input is guaranteed to be deterministic.



In FRJ, the default reference capability is carefully designed to require explicit syntax
to introduce any impurity and non-determinism. Because \Q@imm@ is the default reference capability,
imperative features are controlled by default.
The values of signals are always \Q@imm@, so every other reference being \Q@imm@ by default makes using signals easier. Additionally, outside of the main expression, all classes may not perform any I/O or other
side effects without being declared as a \Q@capability@ or taking an object capability as input.

\subsection{Typing Rules}


FRJ's typing environment has three components:
$\Gamma$, the mapping between variables and types;
$\M$, the mapping between a memory address and object locations;
and $\Cap$, a flag identifying if the expression is allowed to instantiate capability classes.

We will use notation $\capOf(\C)$ and $\capOf(\T)$ to denote the capability modifier of a given class.

\[
    \trfrac[(\x)]{
    }{
        \envs \x : \Gamma(\x)
    }
\]

\[
    \trfrac[(sub)]{
        \envs \e : \T'\qquad
        \vdash \T' \leq \T
    }{
        \envs \e : \T
    }
\]


\[
    \trfrac[(\L)]{
    }{
        \envs \L : \mdf\, \M(\L)
    }
\]

Variable typing and subsumption are standard.

The rule (L) types memory references as the class of the object it points to and the modifier of the reference\footnote{To complete a proof of soundness, we would likely need to instrument the reduction to keep track of the pair $\L{:}\mdf$.}.

\[
    \trfrac[(fAccess)]{
        \envs \e : \mdf\, \C\quad
        \T_1\, \f_1\ldots\T_n\, \f_n = \fields(\C)
    }{
        \envs \e\sdot\f_i : \T_i + \mdf
    }
\]

\[
    \trfrac[(fUpdate)]{
        \begin{trgather}
            \envs \e_1 : \,\mdfMut\, \C\quad
            \T_1\, \f_1\ldots\T_n\, \f_n = \fields(\C)\\
            \envs \e_2 : \T_i
        \end{trgather}
    }{
        \envs \e_1\sdot\f_i \Q@=@ \e_2 : \T_i
    }
\]

Field access and field update are conventional with the exception of modifiers being applied to
the result of a field access and the added requirement that the receiver of a field update must be
\,\mdfMut\,.
The rules for the composition for the reference capabilities of the result of a field access are:
\begin{itemize}
    \item $\overline{@}\ \mdf\ \C + \,\mdfImm = \overline{@}\ \mdfImm\,\ \C$
    \item $\overline{@}\ \mdf\ \C + \,\mdfMut = \overline{@}\ \mdf\ \C$
    \item $\overline{@}\ \mdf\ \C + \,\mdfCapsule = \overline{@}\ \mdf\ \C$
    \item $\overline{@}\ \mdfMut\,\ \C + \,\mdfRead = \overline{@}\ \mdfRead\,\ \C$
    \item $\overline{@}\ \mdfImm\,\ \C + \,\mdfRead = \overline{@}\ \mdfImm\,\ \C$
\end{itemize}
For example, with a field access, if the receiver had the \Q@read@ modifier and the field had
the \Q@imm@ modifier, result would be \Q@imm@. Alternatively, if the receiver was \Q@read@ and the field
was \Q@mut@, the result would be \Q@read@.

\[
    \trfrac[(new)]{
        \begin{trgather}
            \T_1\, \f_1\ldots\T_n\, \f_n = \fields(\C)\quad
            \envs \e_i : \T_i
            \textbf{}\quad
            \text{either } \capOf(\C) = \emptyset \text{ or } \Cap=\capability
        \end{trgather}
    }{
        \envs \new{\e_1\ldots\e_n}\ :\ \mdfMut\ \C
    }
\]

Object instantiation is also mostly conventional. The major difference is that
if the class is marked as \capability, then the object can only be created in
the main method or in a \Q@mut@ method of another capability class ; see rule (method) on page 11.

\[
    \trfrac[(newImm)]{
        \begin{trgather}
            \T_1\, \f_1\ldots\T_n\, \f_n = \fields(\C)\quad
            \envs \e_i : \T_i[\mdf=\mdfImm\,]
        \end{trgather}
    }{
        \envs \new{\e_1\ldots\e_n}\ : \,\mdfImm\,\ \C
    }
\]

If the constructor arguments are all \,\mdfImm\,, then the object created
can be typed with the \,\mdfImm\, modifier; also capability classes can be instantiated by this rule, since only the \Q@mut@ methods can do privileged operations.

\[
    \trfrac[(mCall)]{
        \begin{trgather}
            \T_0\ldots\T_n \mapsto \T\ \text{in}\ \methTypes(\T_0, \m)\quad
            \envs \e_i : \T_i 
        \end{trgather}
    }{
        \envs \e_0\,\mcall{\e_1\ldots\e_n}\ : \T
    }
\]

Our method call type rule is mostly conventional but relies on $\methTypes$, and thus is more flexible than the conventional one.
\[
    \trfrac[(mCall@)]{
        \begin{trgather}
            \envs \e_0 : \T_0\qquad
            T_0\ldots\T_n \mapsto \T\ \text{in}\ \methTypes(\T_0, \m)\\
            \envs \e_i : \at\T_i\ \ \forall i\in1..n\qquad
            \validActor(\T_0)
        \end{trgather}
    }{
        \envs \e_0\,\lmcall{\e_1\ldots\e_n}\ : \at\T
    }
\]

The major difference between rule (mCall) and rule (mCall@) is that all of the argument types are lifted ($\at\T$) and the receiver must be a $\validActor(\T_0)$: either the receiver is immutable ($\T_0= \mdfImm\ \_$) or the receiver is a capability instance and have only \Q@imm@ fields ($\capOf(\T_0)=\capability$ and $\mdfMut\, \C\, \f \not\in \fields(\T_0)$).

Actors may receive messages in any order; while 
immutable actors cannot be influenced by such order, a mutable actor may use the messages to update the value of a
field\footnote{If such an actor could be freely created, then we could use it to forge a no-args method with a nondeterministic result.}.

$\validActor(\T_0)$ prevents this issue, but it requires mutable actors to be instances of capability classes. Note that there is no need for all of the actors to be created in \Q@main@; it is sufficient to create a single capability \Q@ActorSystem@ object that creates new actors using some \Q@mut@ method.

\[
    \trfrac[(fullSignal)]{
        \begin{trgather}
            \Cap;\M;\Gamma[\mathrm{only}\ \mdfImm\,,\ \mdfCapsule\,] \vdash \e_1 : \T\qquad
            \Cap;\M;\Gamma[\mathrm{only}\ \mdfImm\,,\ \mdfCapsule\,] \vdash \e_2 : \at\T
        \end{trgather}
    }{
        \envs \scon{\at}{\e_1}{\e_2} : \at\T
    }
\]

The rule (fullSignal) is for a signal constructor
with both a head and a tail.
The rule enforces that only \,\mdfImm\, and \,\mdfCapsule\,
variables can be captured by the deferred executed expressions inside the signal.

\[
    \trfrac[(emptySignal)]{
    }{
        \envs \sconE : \at\T
    }
\]

The rule (emptySignal) is similar to the conventional rule for typing empty lists, as the empty signal can assume any signal type; not unlike how \Q@[]@ in Haskell is generic and valid for any list type.

\[
    \trfrac[(head)]{
        \envs \e : \at\T
    }{
        \envs \head{\e} : \T
    }
\]

\[
    \trfrac[(tail)]{
        \envs \e : \at\T
    }{
        \envs \tail{\e} : \at\T
    }
\]

Rule (head) extracts the type of the value in the head
and rule (tail) preserves the type of the expression.

\[
    \trfrac[(class)]{
        \begin{trgather}
            \Cap; \C \vdash \Meth_i\qquad
            \overrideOk(\C',\Meth_i)
            \quad\forall \C' \in \Cs\\
            \mathrm{dom}(\C')\subseteq\mathit{dom}(\C)\quad\forall \C' \in \Cs
        \end{trgather}
    }{
        \vdash \Cap\ \ \class\ \C\ \Q@implements@\ \Cs\ \oC\Fs\ \K\ \Meth_1\ldots\Meth_n \cC\ \OK
    }
\]

\[
    \trfrac[(interface)]{
        \begin{trgather}
    \overrideOk(\C',\MH_i)
        \quad\forall \C' \in \Cs\ 
        \end{trgather}
    }{
        \vdash\Q@interface@\ \C\ \Q@extends@\ \Cs \oC\!\MH_1\ldots\MH_n\!\cC\ \OK
    }
\]

\[
    \trfrac[(method)]{
        \begin{trgather}
            \Cap';\,\this\, : \mdf\ \C, \x_1:\T_1\ldots\x_n:\T_n \vdash \e : \T\\
            \Cap' =\emptyset \text{ iff } \Cap=\emptyset \text{ or } \mdf \neq \,\mdfMut
        \end{trgather}
    }{
        \Cap, \C \vdash \mdf\ \method\ \T\ \m\oR\T_1\,\x_1\ldots\T_n\,\x_n \cR\oC\return\ \e\,\semi\cC
    }
\]

The last three type rules (class, interface, and method) are standard
with the exception of rule (method), where every \,\mdfMut\, method in a
\capability\ class is typed as a capability method.
We omit the trivial but tedious definition for 
$\overrideOk(\C',\MH_i)$, checking if a method signature can override a potential method with the same name defined in the super interface: if another method with the same name exists, the two method types must be identical.

\section{Example}
The scenario used in the proposal of the first-order purely FRP language, \textit{Emfrp} \cite{watanabe_sawada_2016}, is an air conditioning
unit's controller. The inputs are \textit{temperature}, \textit{humidity}, and the \textit{current power
state} of the unit. The output is what the power state of the unit should be. To show how FRJ works, the same scenario
can be done with our system. For this example we are taking the liberty of using number/boolean literals and infix operators
for simplicity's sake.

\!\!\!We assume the existence of two capability classes: \Q@Sensors@,
which contains methods to read the physical sensors on the
AC unit and a clock; and \Q@AC@, which interacts directly with the hardware to change power states.

A \textit{discomfort index} is calculated based on the temperature and the humidity to determine
how uncomfortable the room is. We represent that with this actor:
\begin{lstlisting}
class ComfortComputer {
  method Float discomfort(Float temp,Float hum){
    return 0.81 * temp + 0.01 * hum * (0.99 * temp - 14.3) + 46.3; 
    }
  }
\end{lstlisting}
The actor \Q@ACController@ computes if the unit should be on or off, depending on
the discomfort index and its current power state. The current power state of the unit is needed
to apply hysteresis, so that the unit does not constantly change power state. 
The first implementation is in the traditional actor style with mutable state:
\begin{lstlisting}
capability class ACController {
  Bool isOn; //can be updated
  ACController(Bool isOn){ this.isOn = isOn; }
  read method Float hysteresis() {
    return this.isOn?-0.5:0.5;
    }
  mut method Bool powerSwitch(Float d){
    this.isOn = d >= 75.0 + this.hysteresis();
    return this.isOn;
    }
  }
\end{lstlisting}
Note how \Q@ACController@ is a valid mutable actor:
It is a capability class where the only field is of type \Q@imm Bool@. Note how the field can still be updated (line 7); FRJ only requires the referred object (the \Q@Bool@) to be deeply immutable.

We now have an actor that will generate current discomfort values and another actor that will
determine the current power state. In \Q@main@, we can connect them to the sensors to make our
program react to the real world:
\begin{lstlisting}
//Get sensor input
mut Sensors s = new Sensors(); //capability
@Bool tick = s.clock(); //emits every second
@Float temps = s.@temp(tick);
@Float humidities = s.@humidity(tick);
//Decide power state
@Float discomfort = new ComfortComputer().@discomfort(temps,humidities);
@Bool powerState = new ACController(false).@powerSwitch(discomfort);
//Apply power state
mut AC ac = new AC(); //capability
ac.@setPower(powerState);
\end{lstlisting}

\noindent One nice feature of our system is that, because inputs are waited for, the \Q@tick@ input coordinates the
system to update once a second at maximum. The \Q@tick@ dependency is similar to using Rx's \Q@Interval@ operator
as a source \cite{rxInterval}.
That feature is important: it avoids mailbox overflow when one
sensor is faster than the other.

We now reimplement \Q@ACController@
in a more traditional FRP approach, 
where state is kept
via recursion with signals, much like Elm's \Q@foldp@ pattern (before Elm removed FRP from their language\footnote{\url{https://elm-lang.org/news/farewell-to-frp}})
\Q@foldp@ is short for \emph{``fold over the past''}\cite{czaplicki2013elm} and is typed
\Q|(a -> b -> b) -> b -> @a -> @b|.
We can define \Q@FoldP@ in FRJ, extended with some modern Java features:

\begin{lstlisting}
interface FoldP<I, O> {
  method O apply(I v, O old);
  static method <I,O> @O of(FoldP<I,O> f, O initial, @I signal){
    O out = f.apply(head(signal), initial);
    return @[out; FoldP.of(f, out, tail(signal))];
    }
  }
class ACController{//functional using FoldP
  method Float hysteresis(Bool isOn){ return isOn?-0.5:0.5; }
  method @Bool powerSwitch(@Float discomfort){
    return FoldP.of(
      (d, isOn)-> d >= 75.0 + this.hysteresis(isOn),
      false,
      discomfort);
    }
  }
\end{lstlisting}

To switch to the FRP style while keeping the same behaviour,
on line 7 of the previous main we can use \\* \Q{new ACController().powerSwitch(discomfort)}.

Both programming styles are highly parallelisable, with clear dependency chains,
and a fairly compact code footprint.
FRJ allows for smooth transitions between the FRP and Actor model
approaches to concurrent programming.

\section{Related Work}
The potential for connection between reactive programming and the actor model has been a
subject of active research over the past 4 years. The reactor model's attempt
to create deterministic actors using reactive programming \cite{determActors} offers guarantees on deterministic
message delivery and processing order to ensure that all nodes requesting an input get it and process it before
the next message is sent.
However, by abstracting message passing with FRP's \textit{signal} primitive, FRJ enables immutable and pure
actors that do not need to be bound by the reactor model's strict ordering rules.

Work has been done by Van den Vonder et al. \cite{van2017awkward} on the `actor-reactor model' (ARM), which instead
of replacing actors with reactors, creates a joint model, where actors can be nondeterministic.
The reactors in the ARM should not be confused with the Lohstroh et al.'s reactors; ARM's reactors are always pure and deterministic.
The ARM approach is novel and sensible, but FRJ takes a different path. FRJ does not have the distinction between `actors' and `reactors'.
Instead, we attempted to unify the two systems. FRJ's unified approach does still make a distinction between deterministic and nondeterministic
actors using object capabilities, but a pure FRJ actor is able to perform more complex tasks than an ARM reactor. Ultimately, the ARM is very
compelling, but we think that a unified approach results in simpler systems.

XFRP \cite{shibanai2018distributed} offers an interesting model for executing pure FRP on an actor-based runtime. Using XFRP would be a similar
experience to using FRJ without any object or reference capabilities. The language has fewer sources of non-determinism to control because
it delegates side effects to components that are external to the program. Shibani et al. note their main source
of nondeterminism as the \Q|@last| operator, which is essentially syntactic sugar for \Q@foldp@. Glitch is a common issue with systems inspired by FRP. Single source glitch freedom means that all nodes (lifted functions for FRJ) that have one signal
as an input, will get updates at the same time \cite{myter2019distributed}. If a system does not have glitch freedom,
then parts of the application that depend on the same signals
could be in an inconsistent state until they get the latest message.
If multiple stateful inputs are given to
a signal function, glitch freedom can be violated in XFRP. XFRP manages to get around the issue by adding an option to change the semantics
of their language to the same as FRJ's lifted method call, with a feature they call \textit{source unification}.
FRJ effectively treats all arguments to a lifted function as a single input. FRJ's behaviour
has some interesting implications for glitch freedom. Because FRJ's evaluation model provides for single source
glitch freedom in the same way as XFRP \cite{shibanai2018distributed},
and all arguments to a lifted function can be considered a single source;
FRJ has complete glitch freedom \cite{margara2018consistency}.

As a possible implementation technique, FRJ actors can be implemented with a variety of actor frameworks,
including Akka \cite{akka}. Alternatively, there is no reason why a simpler technique, like Emfrp's actor
implementation \cite{watanabe_sawada_2016} could not be used.
Similarly to \textit{Pony} \cite{clebsch2015deny}, our actor system uses shared-memory message passing guarded by reference and object capabilities.

\section{Conclusions and Future Work}
In this foundational work, we defined FRJ, a core OO calculus modelling both FRP and actor systems. FRJ supports traditional imperative field updates and I/O, but it keeps control of side effects using reference and object capabilities.
The work on FRJ is far from complete, we plan to formally model generics and lambdas, and to study possible efficient highelyimplementation strategies. Garbage collection may require particular attention since it can stop running computations.
We plan to relax the restriction on the state of mutable actors, and to develop some case study, to explore useful programming patterns mixing Actors and FRP,
and potentially to look into applying more performant and newer forms of FRP such as \textit{Yampa}'s version of arrowized FRP \cite{chupin2019yampa, nilsson2002arrows}.
FRJ can model most Actor, FRP, and RP patterns. For example, a signal supplier can model hot or cold signals \cite{liberty2011programming} by either returning a reference to an existing signal or by returning a newly created one.
FRJ's signals can be finite or infinite, and they can either be connected with real world devices or just manipulate objects in memory. FRJ streams can be dynamically created and wired while preserving equational reasoning for all expressions that only take in immutable values as input.

We constructed a proof-of-concept compiler\footnote{\href{https://github.com/NickGeek/frj-compiler}{https://github.com/NickGeek/frj-compiler}}, which helped aid in our confidence of the correctness of the formal model. Many test programs, including the air conditioning examples, were written and compiled to evaluate FRJ. With the sensor delay disabled to maximise throughput, the actor-style example
processed an average of 2,070,425 events per second (0dp) on an Intel Core i7-9750H. The more FRP-native style with signals keeping state via recursion performed roughly equivalently to
the actor style, with an average of 2,072,078 events per second. The compiled programs themselves are highly parallelised, with the air conditioning examples making use of all twelve virtual cores
on the test machine.

Although a proof is still future work, we believe FRJ preserves two formal properties:
\begin{itemize}
    \item If the reduction of an expression $e$ is nondeterministic, then $e$ refers to a pre-existing mutable value.
    \item There are no data races, that is: for all well-typed expressions, two different nondeterministic reduction steps will not execute a field update on the same receiver object.
\end{itemize}

\bibliographystyle{splncs04}
{ \small \bibliography{wflp2020}}

\end{document}